\documentclass[aps,prb,showpacs,twocolumn,floatfix,amssymb]{revtex4-1} 

\usepackage{graphicx}
\usepackage[intlimits]{amsmath}
\usepackage{color}

\allowdisplaybreaks[4]

\begin{document}

\title{Hall and dissipative viscosity effects on edge magnetoplasmons}
\author{Roie Cohen}
\author{Moshe Goldstein}
\affiliation{Raymond and Beverly Sackler School of Physics and Astronomy, Tel Aviv University, Tel Aviv 6997801, Israel}

\begin{abstract}
Hydrodynamic and viscous effects in electronic liquids are at the focus of much current research. Most intriguing is perhaps the non-dissipative Hall viscosity, which, due to its symmetry-protected topological nature, can help identify complex topological orders. In this work we study the effects of viscosity in general, and Hall viscosity in particular, on the dispersion relation of edge magnetoplasmons in 2D electronic systems. Using an extension of the standard Wiener-Hopf technique we derive a general solution to the problem, accounting for the long-range Coulomb potential. Among other features we find that on the edge viscosity affects the dispersion already to leading order in the wavevector, making edge modes better suited for its measurement as compared to their bulk counterparts.
\end{abstract}

\maketitle

\section{Introduction} \label{sec:introduction}
Recently hydrodynamic electron flow has been at the focus of much research effort, both theoretical and experimental~\cite{gurzhi68,dejong95,mueller09,andreev11,titov13,davidson14,briskot15,torre15,ghahari16,alekseev16,levitov16,bandurin16,lucas16b,moll16,crossno16,kumar17,scaffidi2017hydrodynamic,gooth17,narozhny17,kashuba18,lucas18,levin18,kiselev18,berdyugin18}.
In the hydrodynamic regime electronic dynamics is dominated by viscosity, rather than impurity scattering.
Denoting the electronic stress tensor by $T_{ij}$ and its strain by $u_{ij}$, one generally has~\cite{landau6}
\begin{align}
{T_{ij}}=-\sum_{kl} {{C}_{ijkl}}{{u}_{kl}} + {{\eta }_{ijkl}}{\partial_t {u}_{kl}}, \label{eq:stress}
\end{align}
where $C_{ijkl}$ is the tensor of elastic moduli, which reduces to the compressibility for a fluid, and $\eta_{ijkl}$ is the viscosity tensor.
Viscosity is usually associated with energy dissipation. However, this is not always the case: The viscosity tensor $\eta_{ijkl}$ can be decomposed into its symmetric and antisymmetric part (with respect to exchanging the first and last pairs of indexes),
\begin{align}
  \eta_{ijkl} = \eta^+_{ijkl} + \eta^-_{ijkl}, \label{eq:viscosity}
\end{align}
with $\eta^{\pm}_{ijkl} = \pm \eta^{\pm}_{klij}$.
The symmetric part, $\eta^{+}_{ijkl}$, is even under time reversal and indeed causes energy dissipation, but the antisymmetric part, $\eta^{-}_{ijkl}$, is odd under time reversal and therefore dissipationless. The latter part may thus arise when time reversal symmetry is broken (either explicitly, by, e.g., applying an external magnetic field, or spontaneously), and is termed ``Hall viscosity'', in analogy to Hall conductivity/resistance, which is also nondissipative~\cite{chapman,marshall,steinberg1958viscosity,kaufman1960plasma,thompson1961dynamics,braginskii1965transport,avron95,avron98,levay95,tokatly07,tokatly09,read09,haldane09,read11,hughes11,barkeshli12,hoyos12,bradlyn12,wiegmann13,can14,abanov13,gromov14,gromov15}.

In 2D isotropic systems the symmetric viscosity has two independent components, the bulk viscosity $\zeta$ and the shear viscosity $\eta_s$,
\begin{align}
{\eta }^{+}_{ijkl}=
\zeta {{\delta }_{ij}}{{\delta }_{kl}}+
{{\eta }_{s}}\left( {{\delta }_{ik }}{{\delta }_{jl}}+{{\delta }_{il}}{{\delta }_{jk}}-{{\delta }_{ij}}{{\delta }_{kl}} \right), \label{eq:etap}
\end{align}
whereas the Hall viscosity has a single independent component, $\eta_H$,~\cite{avron98}
\begin{align}
{\eta }^{-}_{ijkl}={{\eta }_{H}}\left( {{\delta }_{jk}}{{\epsilon }_{il}} - {{\delta }_{il}}{{\epsilon }_{kj}} \right). \label{eq:etam}
\end{align}
Hall viscosity can exist in a 2D electron fluid in the presence of a strong magnetic field in the quantum Hall regime~\cite{ezawa}, where dissipative effects are suppressed at low temperatures. 
Furthermore, in quantum Hall systems $\eta_H$ is expected to be quantized (assuming rotational invariance) and equal to half the particle density times the orbital spin $\bar{s}$ (one half of the shift quantum number)~\cite{read09,read11,bradlyn12}. Hence, measuring the value of Hall viscosity in an experiment can help one distinguish between different theoretical proposals concerning the nature of quantum Hall states in certain fractional filling factors. Since some of these states have non-Abelian anyonic excitations (with potential applications for topological quantum computing~\cite{nayak08}), knowing the value of Hall viscosity can help in their identification. 
However, and despite recent progress in understanding the relation between viscosity and charge transport in the presence of non-uniform electric fields~\cite{hoyos12,bradlyn12,hoyos14,moroz15,holler16,sherafati16,pellegrino17,ganeshan17,delacretaz17}, so far its measurement has only been possible for very weak magnetic fields, far from the quantum Hall regime~\cite{berdyugin18}. 

The elementary excitations of an ordinary fluid are sound waves; in charged fluid (such as an electron fluid) they become plasmons, or magnetoplasmons in the presence of a magnetic field. Bulk magentoplasmons are gapped excitations, and in the long wavelength limit, the magnetoplasmon frequency approaches a finite value, which is determined by the cyclotron and plasma frequencies. In addition, there are also gapless edge magnetoplasmons (magnetoplasomons confined to the edge of the system) whose frequency vanishes as their wave vector goes to zero. In the quantum Hall regime the quantized version of these excitations give rise to the edge modes which determine the low-energy transport properties of these systems. This allows one to extract their properties in an experiment~\cite{glattli85,fetter85,volkov85,volkov86,fetter86a,fetter86b,nazin88,volkov88,andrei88,grodnensky90,wassermeier90,ashoori92,talyanskii93,zhitenev93,aleiner94,aleiner95,sukhodub04,kamata10,kumada11,petkovic13,sasaki16,bosco17,endo18,jin18}.
Recently analogous excitations induced by Berry curvature rather than magnetic field have attracted significant attention~\cite{song16,jin16,principi16,principi17,zhang17,jia17,mahoney17,zhang18}.

In this work we study the effects of Hall viscosity on the edge magnetoplasmon dispersion (related questions for neutral fluids were recently considered in Refs.~\onlinecite{abanov18,souslov18}). We use a classical hydrodynamic model for a two dimensional electrons fluid with an edge. In contrast to most works in the field, we account for the long-range Coulomb interaction and the resulting logarithmically-divergent (at long wavelength) mode velocity \cite{volkov88,aleiner94,aleiner95} instead of using the Fetter approximation \cite{fetter85}; To the best of our knowledge, the interplay between the long-range Coulomb potential and viscosity has not been considered before. For this we develop a modification of the standard Wiener-Hopf technique~\cite{stone}. We find that whereas in the bulk the viscosity only affects the dispersion through high-order terms in the wavevector, on the edge it modifies the leading order term, making it easier to probe.
In Sec.~\ref{sec:model} we will introduce our hydrodynamic model, and then present our method and general solution for different boundary conditions in Sec.~\ref{sec:method}. Then we will discuss realistic parameter values and show examples of the resulting dispersions in Sec.~\ref{sec:results}, followed by the conclusions, Sec.~\ref{sec:conclusions}. Some technical details are relegated to the Appendixes.
   
\section{Model} \label{sec:model}
\begin{figure} 
	\centering
	\includegraphics[width=\columnwidth]{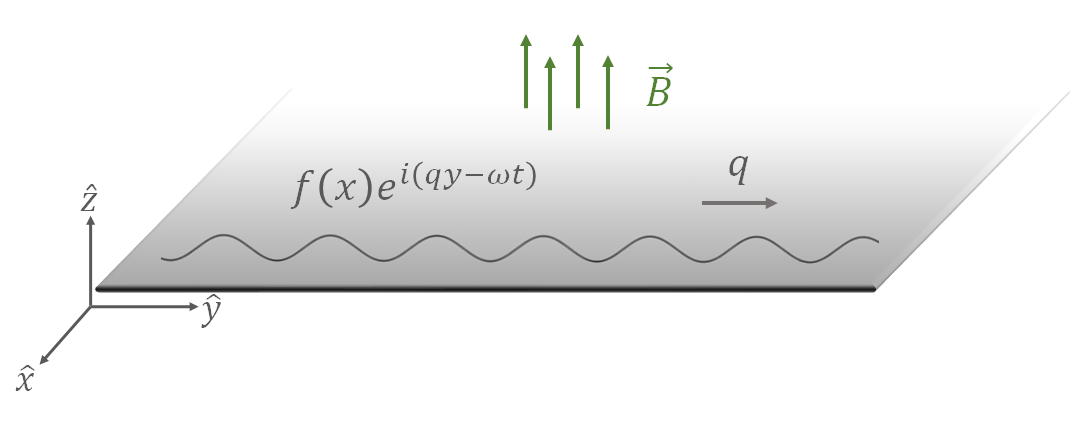}
	\caption{A schematic depiction of the system: a 2D electronic system with an edge under the influence of a perpendicular magnetic field. See the text for further details.}
	\label{fig:system}
\end{figure}
We consider a 2D system on a half infinite plane with an edge (see Fig.~\ref{fig:system}), occupying the region $x<0$, with a magnetic field $\mathbf{B}=B\hat{z}$ perpendicular to the plane. We employ a linearized hydrodynamic model of a weakly damped, compressible, viscous fluid. The state of the system is characterized by the electron number density $n=n_0+\delta n$ ($n_0$ being the equilibrium density and $\delta n$ a small deviation), and the electron velocity $\mathbf{v}$. The interaction between the electrons is taken into account through the electric potential $\phi$.
These quantities are related through the 2D linearized continuity equation,
\begin{align}
\partial_t \delta n + {{n}_{0}}\mathbf{\nabla } \cdot \mathbf{v}=0, \label{eq: continuity}
\end{align}
the 2D linearized Navier-Stokes equation,
\begin{align}
\label{eq: NS}
\partial_t \mathbf{v} + \frac{\mathbf{v}}{\tau }=&
-\frac{{{s}^{2}}}{{{n}_{0}}}\mathbf{\nabla }\delta n+\frac{e}{m}\mathbf{\nabla }\phi +{{\omega }_{c}}\hat{z} \times \mathbf{v}+
\\
&\frac{1}{{{n}_{0}}m}\left( {{\eta }_{s}}{{\nabla }^{2}}\mathbf{v}+ \zeta \mathbf{\nabla } \left( \mathbf{\nabla }\cdot\mathbf{v} \right)-{{\eta }_{H}}{{\nabla }^{2}}\left( \widehat{z}\times \mathbf{v} \right) \right), \nonumber 
\end{align}
and the Poisson equation,
\begin{align}
{{\nabla }^{2}}\phi =\frac{4\pi e}{\epsilon} \delta n\Theta (-x)\delta (z) , \label{eq: Poisson}
\end{align}
where $\tau$ is the electronic momentum relaxation time due to collisions with impurities, $\omega_c=eB/mc$ is the cyclotron frequency ($-e$ and $m$ are the electron charge and \emph{effective mass}, respectively, and $c$ is the speed of light), and $\epsilon$ is the dielectric constant of the 3D medium. 
In addition, $\Theta$ is the Heaviside step function and $\delta$ is the Dirac delta function. 
Finally, $s$ is the sound velocity in the absence of the Coulomb interaction, $s^2 = (d P/dn)/m$, where $P$ is the noninteracting internal pressure.
We note that the internal pressure is the equilibrium average of the diagonal elements of the stress tensor. It is different from the pressure on the boundaries, which contains the effects of the Lorentz force on the boundary currents. While the compressibility associated with the pressure on the boundaries vanishes in the quantum Hall regime, the internal pressure is associated with a finite internal compressibility, hence with a finite $s^2$.~\cite{bradlyn12,cooper1997thermoelectric}. Thus, our model should apply to the quantum Hall regime, provided one plugs into it the appropriate (and generally frequency dependent) viscosities etc.

As for boundary conditions, one of them is straightforward: The vanishing of the normal velocity at the edge, $v_x(0,y)=0$. The other one is more subtle~\cite{kiselev18}. To simplify the discussion we will start from taking the no-slip boundary condition, $v_y(0,y)=0$. Later on we will show how the results are modified for the no-stress boundary condition (zero tangential force on the boundary), $T_{xy}(0,y)=0$.
As we will further show below, our method allows one to treat more complicated electrostatic environment of the 2D system, including multiple dielectrics and metallic gates.

\section{General Solution} \label{sec:method}

Since the system is invariant under translations in the $y$ direction, we can look for a solution which is a plane wave parallel to the edge, $\propto {{e}^{iqy}}$. In the following we do not write explicitly the q argument of the physical quantities. Fourier-transforming in the $x$ direction we find
\begin{align}
\left[ \frac{{{d}^{2}}}{d{{z}^{2}}}-\left( {{k}^{2}}+{{q}^{2}} \right) \right]\bar{\phi }(k,z)=\frac{4\pi e}{\epsilon} {{\overline{\delta n}}_{+}}(k)\delta (z) , \label{eq: Poisson transform}
\end{align}
where $\bar{\phi}(k,z)\equiv \int_{-\infty }^{\infty}{\phi(x,z){{e}^{-ikx}}dx}$ and $\overline{\delta n}_+(k)\equiv \int_{-\infty }^{0}{\delta n(x){{e}^{-ikx}}dx}$; similar notations will be used for the $x<0$ Fourier transform of other physical quantities, which are analytic in the upper complex $k$-plane (provided the physical quantity does not diverge exponentially at large $x$).
The solution is then $\bar{\phi}(k,z) =\bar{\phi}(k,0) e^{-|z|\left(k^2+q^2\right)^{1/2}}$, where in the system plane $z=0$:
\begin{align} 
\bar{\phi }(k,0) = -\frac{4\pi e}{\epsilon} \bar{L}(k) \overline{\delta n}(k), \label{eq: potential Fourier} 
\end{align}
with the Poisson kernel $\bar{L}(k) \equiv1/2\sqrt{{{k}^{2}}+{{q}^{2}}}$. In the following calculations we will not rely on the specific form of $\bar{L}(k)$. This will make our results applicable to more complicated geometries (with dielectric media, gates, etc.), whose effect can be accounted for by a modified $\bar{L}(k)$.

In the bulk one may combine this results with the Fourier-transformed continuity equation~(\ref{eq: continuity}) and Navier-Stokes equation~(\ref{eq: NS}), and look for a nontrivial solution for the resulting homogeneous linear algebraic equations for the Fourier coefficients. One may thus straightforwardly obtain an implicit equation for the dispersion relation $\omega(\mathbf{p})$, where $\mathbf{p}=(k,q)$,
\begin{align}
\frac{\omega}{\omega_s} \left(\omega_s^2 - \omega^2_H \right)=
\Omega_p^2 + s_\zeta^2 p^2, \label{eq: bulk}
\end{align}
where 
$\Omega_p = \sqrt{2\pi n_0 e^2 p/m\epsilon}$ is the 2D plasma velocity,
${{\omega}_{H}}\equiv {{\omega }_{c}}+{{\gamma }_{H}} p^2$,
${{\omega }_{s}}\equiv \tilde{\omega} + i{{\gamma }_{s}} p^2$ ($\tilde{\omega}\equiv\omega+i/\tau$),
and ${{s}_{\zeta }}^{2}\equiv {{s}^{2}}-i{{\gamma }_{\zeta }}\omega $, with
${{\gamma }_{a}} \equiv {{\eta }_{a}}/{{n}_{0}}m$ for $a=H,s,\zeta$.
One thus sees that the bulk dispersion is affected by the viscosity only to second order in $q$.~\cite{tokatly07,tokatly09}

The edge problem is much more involved. To simplify it the Fetter approximation~\cite{fetter85} is often used (Appendix~\ref{sec:fetter}), the Coulomb kernel $\bar{L}(k)$ is replaced by a meromorphic function, allowing one to get an effective 2D ``Poisson'' equation for the potential. 
However, this approximation qualitatively affects the solution; most notably, it gives a finite group velocity at long wavelengths, as opposed to the true logarithmic divergence in the velocity, due to the long range Coulomb interaction~\cite{volkov88}.
We will therefore refrain from such approximations, and instead extend the Wiener-Hopf method~\cite{volkov88,stone} to the current system. We will later on use the Fetter approximation as a test case for our general solution, comparing the direct solution it allows with the one obtained from our general solution when $\bar{L}(k)$ is replaced by Fetter's approximate kernel (Appendix~\ref{sec:fetter}).

\subsection{No-slip boundary conditions}
Let us start with the no-slip boundary conditions.
Taking the Fourier transform of Eqs.~(\ref{eq: continuity})--(\ref{eq: NS}) for $x<0$, and using the boundary conditions leads to
\begin{widetext}	
\begin{subequations} 
\begin{align}
\begin{split}
-i{{\omega }_{s}}\bar{v}_{x}^{+}+\frac{{s}_{\zeta }^{2}}{{{n}_{0}}}\left( ik{{{\overline{\delta n}}}_{+}}+\delta n(0) \right)-\frac{e}{m}\left( ik{{{\bar{\phi }}}_{+}}+\phi (0) \right)+ {{\omega }_{H}}\bar{v}_{y}^{+}- {{\gamma }_{H}}{{\partial }_{x}}{{v}_{y}}(0)-{{\gamma }_{s}}{{\partial }_{x}}{{v}_{x}}(0) & =0,
\end{split} \\
\begin{split}
 -i{{\omega }_{s}}\bar{v}_{y}^{+}+iq{s}_{\zeta }^{2}\frac{{{{\overline{\delta n}}}_{+}}}{{{n}_{0}}}-iq\frac{e}{m}{{\bar{\phi }}_{+}}-{{\omega }_{H}}\bar{v}_{x}^{+}+{{\gamma }_{H}}{{\partial }_{x}}{{v}_{x}}(0)-{{\gamma }_{s}}{{\partial }_{x}}{{v}_{y}}(0) & =0,
\end{split} \\
\begin{split}
-i\omega {{\overline{\delta n}}_{+}}+{{n}_{0}}\left( ik\bar{v}_{x}^{+}+iq\bar{v}_{y}^{+} \right) & =0.
\end{split}
\end{align}
\label{eq set: positive Fouriers}
\end{subequations}
\end{widetext}
Combining these equations with Eq.~(\ref{eq: potential Fourier}), and using $\bar{\phi(k,0)} = \bar{\phi}_+(k,0) + \bar{\phi}_-(k,0)$, where $\bar{\phi}_{-}(k,0) \equiv \int_{0}^{\infty}{\phi(x,0){{e}^{-ikx}}dx}$ is analytic in the lower half of the complex $k$ plane, we find
\begin{widetext}
\begin{align}
{{\bar{\phi }}_{-}}+\left( 1+\frac{2{\Omega }_{q}^{2}{{\omega }_{s}}\left( {{k}^{2}}+{{q}^{2}} \right)\bar{L}}{q{s}_{\zeta }^{2}{{\omega }_{s}}\left( {{k}^{2}}+{{q}^{2}} \right)+q\omega \left( {\omega }_{H}^{2}-{\omega }_{s}^{2} \right)} \right){{\bar{\phi }}_{+}}
=
\frac{{s}_{\zeta }^{2}\left( k{{\omega }_{s}}+iq{{\omega }_{H}} \right)\tilde{A}\bar{L}}
{{s}_{\zeta }^{2}{{\omega }_{s}}\left( {{k}^{2}}+{{q}^{2}} \right)+\omega \left( {\omega }_{H}^{2}-{\omega }_{s}^{2} \right)} ,
\label{eq: phi_bar+- relation}
\end{align}
where
\begin{align*}
\tilde{A}(k)=-\frac{4\pi i{{n}_{0}}e}{{s}_{\zeta }^{2}\epsilon}\left( {s}_{\zeta }^{2}\frac{\delta n(0)}{{{n}_{0}}}-\frac{e}{m}\phi (0)+{{\gamma }_{H}}\left( \frac{iq{{\omega }_{s}}+k{{\omega }_{H}}}{ik{{\omega }_{s}}-q{{\omega }_{H}}}{{\partial }_{x}}{{v}_{x}}(0)-{{\partial }_{x}}{{v}_{y}}(0) \right) \right).
\end{align*} 

The coefficient of $\bar{\phi}_+$ can be expressed as the ratio $\bar{X}_-/\bar{X}_+$ of two functions which are analytic and have no zeros in the lower/upper half of the complex $k$ plane, respectively. The solution of this Riemann-Hilbert problem is:
\begin{align}
{{\bar{X}}_{\pm }}(k)=\exp \left\{ -\frac{1}{2\pi i}\int\limits_{-\infty }^{\infty }{\frac{dk^\prime}{k^\prime-k\mp i\epsilon }\ln \left( 1+\frac{2{\Omega }_{q}^{2}{{\omega }_{s}(k^\prime)}\left( k^{\prime 2} + {q}^{2} \right)\bar{L}(k^\prime)}{q{{s}_{\zeta }}^{2}{{\omega }_{s}(k^\prime)}\left( k^{\prime 2}+{{q}^{2}} \right)+q\omega \left( {\omega}^2_{H}(k^\prime)-{\omega}^2_{s}(k^\prime) \right) } \right) } \right\},
\label{eq: X definition}
\end{align}
\end{widetext}
which further obeys
\begin{align*}
\frac{1}{{{\bar{X}}_{+}}}-\frac{1}{{{\bar{X}}_{-}}}
=
\frac{1}{{{\bar{X}}_{-}}}\frac{2{\Omega }_{q}^{2}{{\omega }_{s}}\left( {{k}^{2}}+{{q}^{2}} \right)\bar{L}}
{q{s}_{\zeta }^{2}{{\omega }_{s}}\left( {{k}^{2}}+{{q}^{2}} \right)+q\omega \left( {\omega }_{H}^{2}-{\omega }_{s}^{2} \right)}.
\end{align*}
As a result, Eq.~(\ref{eq: phi_bar+- relation}) can be rewritten as:
\begin{align}
\frac{{{{\bar{\phi }}}_{+}}}{{{\bar{X}}_{+}}}+\frac{{{{\bar{\phi }}}_{-}}}{{{\bar{X}}_{-}}}=\left( \frac{1}{{{\bar{X}}_{+}}}-\frac{1}{{{\bar{X}}_{-}}} \right)\frac{q{s}_{\zeta }^{2}\left( k{{\omega }_{s}}+iq{{\omega }_{H}} \right)\tilde{A}}{2{\Omega }_{q}^{2}{{\omega }_{s}}\left( {{k}^{2}}+{{q}^{2}} \right)} ,
\label{eq: phi_bar+- with X}
\end{align}

The right hand side of the last equation can be expressed as the difference $\bar{\psi}_+(k)-\bar{\psi}_-(k)$ of two functions analytic in the upper and lower $k$ plane, respectively. The solution of this additional Riemann-Hilbert problem is
\begin{align}
{{\bar{\psi} }_{\pm }}(k) =
\frac{1}{2\pi i}\int\limits_{-\infty }^{\infty } & \frac{dk^\prime}{k^\prime-k\mp i\epsilon }
	\left( \frac{1}{{{\bar{X}}_{+}}(k^\prime)}-\frac{1}{{{\bar{X}}_{-}}(k^\prime)} \right)
\times \nonumber \\	
&	\frac{q{s}_{\zeta }^{2}\left( k^\prime{{\omega }_{s}}(k^\prime)+iq{{\omega }_{H}}(k^\prime) \right)\tilde{A}(k^\prime)}{2{\Omega }_{q}^{2}{{\omega }_{s}}(k^\prime)\left( k^{\prime 2}+{{q}^{2}} \right)} .
\label{eq: psi definition}
\end{align}
By rearranging Eq.~(\ref{eq: phi_bar+- with X}) we get the relation:
\begin{align}
\frac{\bar{\phi}_+}{\bar{X}_+} - \bar{\psi}_+ = -\frac{\bar{\phi}_-}{\bar{X}_-} - \bar{\psi}_- .
\label{eq: WH thing}
\end{align}
According to Liouville's theorem, every bounded entire function must be a constant. By our assumptions the functions appearing in Eq.~(\ref{eq: WH thing}) decay to zero at infinity in the complex $k$ plane, so both sides of
Eq.~(\ref{eq: WH thing}) must vanish. This gives us ${{\bar{\phi }}_{+}}={{\bar{X}}_{+}}{{\bar{\psi} }_{+}}$.
Evaluating the integrals in Eq.~(\ref{eq: psi definition}) leads to
\begin{widetext}
\begin{align}
\begin{split}
\frac{e}{m}{{\bar{\phi }}_{+}} = 
-i\left\{ \frac{1}{{{k}^{2}}+{{q}^{2}}}\left[ Ak+iqB+\left( iqA+kB \right)\frac{{{\omega }_{H}}}{{{\omega }_{s}}} \right]+ \right. 
 \left.
\frac{{{\bar{X}}_{+}}}{2\left| q \right|}\left( \frac{{{C}_{-}}}{k+i\left| q \right|}-\frac{{{C}_{+}}}{k-i\left| q \right|} \right)+\frac{{{\bar{X}}_{+}}}{2{{k}_{s}}}\frac{{{\omega }_{H}}({{k}_{s}})}{{\tilde{\omega }}}\left( \frac{{{D}_{+}}}{k-{{k}_{s}}}-\frac{{{D}_{-}}}{k+{{k}_{s}}} \right) \right\},
\end{split}
\label{eq: phi+ explicit}
\end{align}
\end{widetext}
where
\begin{align*}
&A\equiv \left( \frac{{s}_{\zeta }^{2}}{i\omega }-{{\gamma }_{s}} \right){{\partial }_{x}}{{v}_{x}}(0)-\frac{e}{m}\phi (0)-{{\gamma }_{H}}{{\partial }_{x}}{{v}_{y}}(0) , \\ 
&B\equiv i\left( {{\gamma }_{s}}{{\partial }_{x}}{{v}_{y}}(0) - {{\gamma }_{H}}{{\partial }_{x}}{{v}_{x}}(0)\right) , \\
&{{C}_{\pm }}=\frac{{\tilde{\omega }} \left( Bq\pm A\left| q \right| \right) +{\omega }_{c}\left( Aq\pm B\left| q \right| \right)}{{\tilde{\omega }}{{\bar{X}}_{\pm }}(\pm i\left| q \right|)} , \\
&{{D}_{\pm }}=\frac{iqA\pm B{{k}_{s}}}{{{\bar{X}}_{\pm }}(\pm {{k}_{s}})},
\end{align*}
and $k_s$ is the root of
${{\omega }_{s}}({{k}_{s}})=\omega +i/\tau +i{{\gamma }_{s}}\left( {k}_{s}^{2}+{{q}^{2}} \right)=0$
in the upper complex plain.

\begin{figure*} 
	\centering
	\includegraphics[width=\textwidth]{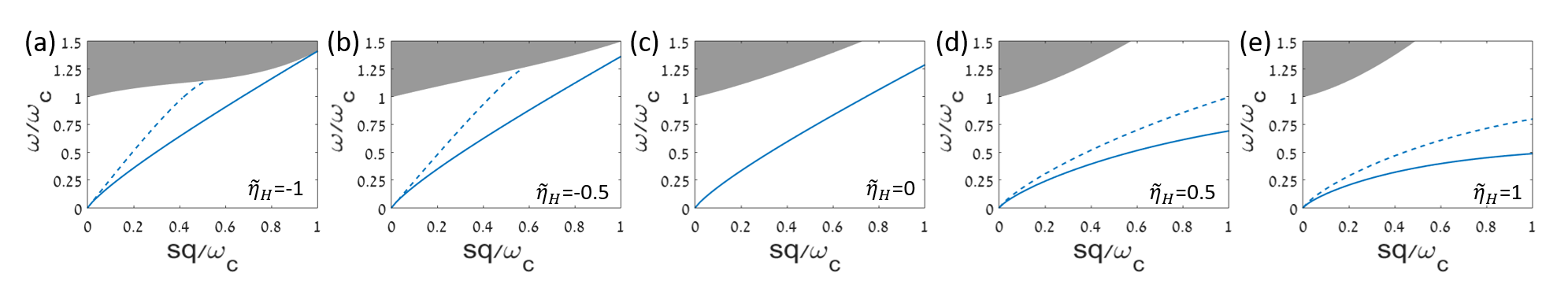}
	\caption{Dissipationless magnetoplasmons: The dispersion relation for the edge magneto-plasmons in the dissipationless limit ($\tau \to \infty$, $\zeta=\eta_s=0$) for $\tilde{\Omega}=1$ and different values of the dimensionless Hall viscosity, $\tilde{\eta}_{H} \equiv \omega_c \eta_H/n_0 m s^2$. The continuous and dashed lines correspond to no-slip and no-stress boundary conditions, respectively. The gray area denotes the bulk magnetoplasmon spectrum.}
	\label{fig:dispersion_dissipationless}
\end{figure*}

Typically in the Wiener-Hopf technique, the next step would be to impose the boundary conditions~\cite{volkov88,stone}. However, in the current case these are satisfied automatically (see Appendix~\ref{sec:consistency}).
Interestingly, as opposed to the common situation, what is not immediately ensured is the analyticity in the appropriate half of the complex $k$ plane 
of the transformed variables. 
It turns out to be sufficient to consider the electron number density (Appendix~\ref{sec:consistency}).
Using Eqs.~(\ref{eq set: positive Fouriers}) we can extract
\begin{align}
\frac{{{{\overline{\delta n}}}_{+}}}{{{n}_{0}}} = \frac{{{\omega }_{s}}\frac{e}{m}\left( {{k}^{2}}+{{q}^{2}} \right){{{\bar{\phi }}}_{+}}+i\left( {{\omega }_{s}}k+iq{{\omega }_{H}} \right)A+i\left( i{{\omega }_{s}}q+k{{\omega }_{H}} \right)B}{{{\omega }_{s}}{s}_{\zeta }^{2}\left( {{k}^{2}}+{{q}^{2}} \right)+\omega \left( {\omega }_{H}^{2}-{\omega }_{s}^{2} \right)} .
\label{eq: n_bar}
\end{align}
It has potential poles in the upper plane for $k$ values for which the denominator in the last equation vanishes, giving
\begin{widetext}
\begin{align}
\begin{split}
\left[ i {\gamma }_{s} {s}_{\zeta }^{2} + \omega \left( {\gamma }_{H}^{2}+ {\gamma }_{s}^{2} \right) \right] {{k}^{4}}
+ 
\left[ {s}_{\zeta }^{2} \left( \tilde{\omega } +2i {\gamma }_{s} {q}^{2} \right)
+2\omega \left( {\omega }_{c}{\gamma }_{H} - i {\gamma }_{s} \tilde{\omega } + \left( {\gamma }_{H}^{2} + {\gamma }_{s}^{2} \right) {q}^{2} \right) \right] {{k}^{2}} +
\tilde{\omega } {s}_{\zeta }^{2} {q}^{2} + i {\gamma }_{s} {s}_{\zeta }^{2} {q}^{4}
+ \omega \left( {\omega }_{H}^{2}-{\omega }_{s}^{2} \right)
=0.
\end{split}
 \label{eq: kh definition}
\end{align}
Let us denote the two roots of this equation in the upper complex $k$ plane as $k_{H1}$ and $k_{H2}$.
The analyticity of $\overline{\delta n}_+(k)$ in the upper half complex $k$ plain amounts to the vanishing of the residues at $k_{H1}$ and $k_{H2}$, giving the two equations:
\begin{align}
\left( {{\omega }_{s}}({{k}_{H}}){{k}_{H}}
+ iq{{\omega }_{H}}({{k}_{H}}) \right)A
+ \left( i{{\omega }_{s}}({{k}_{H}})q+{{k}_{H}}{{\omega }_{H}}({{k}_{H}}) \right)B 
-i\frac{e}{m} {{\omega }_{s}}({{k}_{H}})\left( {k}_{H}^{2}+{q}^{2} \right){{\bar{\phi }}_{+}}({{k}_{H}})
=0 , 
\label{eq: residues vanish}
\end{align}
for $k_H = k_{H1}, k_{H2}$.
Substitution of $\bar{\phi}_+$ to the last equation gives:
\begin{align}
\frac{1}{2\left| q \right|}\left( \frac{{{C}_{-}}}{{{k}_{H}}+i\left| q \right|}-\frac{{{C}_{+}}}{{{k}_{H}}-i\left| q \right|} \right)+\frac{1}{2{{k}_{s}}}\frac{{{\omega }_{H}}({{k}_{s}})}{{\tilde{\omega }}}\left( \frac{{{D}_{+}}}{{{k}_{H}}-{{k}_{s}}}-\frac{{{D}_{-}}}{{{k}_{H}}+{{k}_{s}}} \right)=0.
\label{eq: determinant equations}
\end{align}  
The vanishing of the determinant for these two linear equations (for $k_H = k_{H1}, k_{H2}$) gives the dispersion relation: 
\begin{align}
\begin{split}
  \left[ \right.
  & \left. -{{k}_{s}}\left( {{k}_{H1}}+{{k}_{s}} \right)\left( {{k}_{H1}}-{{k}_{s}} \right)\left( {{k}_{H1}}-iq \right){{\bar{X}}_{+}}(iq)\left( \tilde{\omega }-{{\omega }_{c}} \right)  
 -{{k}_{s}}\left( {{k}_{H1}}+{{k}_{s}} \right)\left( {{k}_{H1}}-{{k}_{s}} \right)\left( {{k}_{H1}}+iq \right){{\bar{X}}_{+}}(-iq)\left( \tilde{\omega }+{{\omega }_{c}} \right) \right. \\ 
 & \left.
 +{{\omega }_{H}}({{k}_{s}})\left( {{k}_{H1}}+iq \right)\left( {{k}_{H1}}-iq \right)\left( {{k}_{H1}}+{{k}_{s}} \right){{\bar{X}}_{+}}(-{{k}_{s}})iq
 -{{\omega }_{H}}({{k}_{s}})\left( {{k}_{H1}}+iq \right)\left( {{k}_{H1}}-iq \right)\left( {{k}_{H1}}-{{k}_{s}} \right){{\bar{X}}_{+}}({{k}_{s}})iq \right] \times \\
 \left[ \right.
 & \left. +\left( {{k}_{H2}}+{{k}_{s}} \right)\left( {{k}_{H2}}-{{k}_{s}} \right)\left( {{k}_{H2}}-iq \right){{\bar{X}}_{+}}(iq)\left( \tilde{\omega }-{{\omega }_{c}} \right)
 -\left( {{k}_{H2}}+{{k}_{s}} \right)\left( {{k}_{H2}}-{{k}_{s}} \right)\left( {{k}_{H2}}+iq \right){{\bar{X}}_{+}}(-iq)\left( \tilde{\omega }+{{\omega }_{c}} \right)
 \right. \\ 
 & \left.
 +{{\omega }_{H}}({{k}_{s}})\left( {{k}_{H2}}+iq \right)\left( {{k}_{H2}}-iq \right)\left( {{k}_{H2}}+{{k}_{s}} \right){{\bar{X}}_{+}}(-{{k}_{s}})
 +{{\omega }_{H}}({{k}_{s}})\left( {{k}_{H2}}+iq \right)\left( {{k}_{H2}}-iq \right)\left( {{k}_{H2}}-{{k}_{s}} \right){{\bar{X}}_{+}}({{k}_{s}}) \right] 
 -\{H1\leftrightarrow H2\}=0.
\end{split} 
 \label{eq: dispertion} 
\end{align}
\end{widetext}
This last equation, together with Eqs.~(\ref{eq: kh definition}) and~(\ref{eq: X definition}), constitute the full solution of the problem.
We have checked that when the Fetter approximation~\cite{fetter85} is substituted into the general solution, it agrees with the direct solution which one can obtain in that case (see Appendix~\ref{sec:fetter}).


\subsection{No-stress boundary conditions} \label{subsec:nostress}
The above derivation assumed the tangential component of the velocity to vanish at the boundary. An alternative commonly-employed boundary condition is the vanishing of the tangential force on the boundary, $T_{xy}(x=0)=0$.~\cite{kiselev18}
From Eqs.~(\ref{eq:stress})-(\ref{eq:etam}) we have
\begin{align}
{{T}_{xy}}=-{{\eta }_{s}}\left( {{\partial }_{x}}{{v}_{y}}+{{\partial }_{y}}{{v}_{x}} \right)+{{\eta }_{H}}\left( {{\partial }_{x}}{{v}_{x}} - {{\partial }_{y}}{{v}_{y}} \right),
\end{align}
hence $T_{xy}(x=0)=0$ gives
\begin{align}
{{v}_{y}}(0)=\frac{1}{iq}\left( {{\partial }_{x}}{{v}_{x}}(0)-\frac{{{\eta }_{s}}}{{{\eta }_{H}}}{{\partial }_{x}}{{v}_{y}}(0) \right).
\label{eq:v0}
\end{align} 

Using this above relation [instead of $v_y(x=0)=0$] and following the steps of the derivation in the previous Subsection, once again we arrive at a pair of equations similar to Eq.~(\ref{eq: determinant equations}) for $k_H = k_{H1}, k_{H2}$. The no-slip case had the simplifying but non-generic property that the three boundary values, $\phi(0)$, $\partial_xv_x(0)$, and $\partial_xv_y(0)$, only entered the solution through the two combinations $A$ and $B$, so Eq.~(\ref{eq: determinant equations}) for $k_H = k_{H1}, k_{H2}$ was sufficient to obtain the dispersion relation. For the no-stress boundary condition this does not happen, and a third equation is required. Such an equation is provided by demanding the boundary value of the potential matches the one obtained from its Fourier transform (a condition which was automatically obeyed in the no-slip case, see Appendix~\ref{sec:consistency})
\begin{widetext}
\begin{align}
\phi (0)=\lim_{\mathrm{Im} k\to \infty} \left[ -ik{{\overline{\phi }}_{+}}(k) \right]=
-\frac{m}{e}\left[ A + B\frac{{{\gamma }_{H}}}{i{{\gamma }_{s}}}
-\frac{\left( {\gamma }_{s}^{2}+{\gamma}_{H}^{2} \right)}{{{\gamma }_{s}}}
\left( {{\partial }_{x}}{{v}_{x}}(0)-\frac{{{\gamma }_{s}}}{{{\gamma }_{H}}}{{\partial }_{x}}{{v}_{y}}(0) \right)
+\frac{1}{2\left| q \right|}\left( {{C}_{-}}-{{C}_{+}} \right)+ 
\frac{{{\omega }_{H}}({{k}_{s}})}{2{{k}_{s}}{\tilde{\omega }}}\left( {{D}_{+}}-{{D}_{-}} \right) \right],
\label{eq:phi0_boundary}
\end{align} 
\end{widetext}
where all the parameters appearing here are as defined previously. This, together with Eq.~(\ref{eq: determinant equations}) for $k_H = k_{H1}, k_{H2}$, is a set of three homogeneous linear algebraic equations for the boundary values $\phi(0)$, $\partial_xv_x(0)$, and $\partial_xv_y(0)$. To ensure a nontrivial solution we equate the determinant of the resulting coefficient to zero, giving us the dispersion relation. 

Let us finally note that one may treat more complicated boundary conditions (intermediate between no-slip and no-stress~\cite{kiselev18}) in a similar manner.

\section{Results} \label{sec:results}

\subsection{Parameters values}

Let us now turn to estimate the values of the parameters appearing in Eqs.~(\ref{eq: continuity})-(\ref{eq: Poisson}).
The electrons density $n_0$ is related to the Fermi wavevector $k_F$ by ${{n}_{0}}={{k}_{F}}^{2}/2\pi$. The number of filled Landau levels $\nu$ is then $\nu = n_0\Phi_0/B$, where  $\Phi_0=hc/e$ is the flux quantum ($h=2\pi \hbar$ being Planck's constant). 
As for the pressure, at low temperatures and without interactions or magnetic fields, the it is given by (the grand canonical energy per unit area):
\begin{align}
P= 
-2 \int^{{{k}_{F}}}{\frac{{{d}^{2}}k}{{{(2\pi )}^{2}}}\left( {{\epsilon }_{k}}-\mu  \right)}=\frac{\pi {{\hbar }^{2}}}{{{m}}}{{n}_{0}}^{2},
\end{align}
The non-interacting $B=0$ sound velocity is then 
\begin{align} \label{eqn:s2_B0}
{{s}^{2}}=
\frac{1}{{m}} \frac{\partial P}{\partial {{n}_{0}}} =
\frac{2\pi {{\hbar }^{2}}}{{{m}}^{2}}{{n}_{0}}.
\end{align}
Let us note that, as discussed above, in the presence of a magnetic field one actually needs to use the internal pressure~\cite{cooper1997thermoelectric}. Using the results of Ref.~\onlinecite{bradlyn12}, we find that in the integer quantum Hall regime,
\begin{align}
{{s}^{2}}=
\frac{1}{{m} n_0} \left(n_0 \frac{\partial P}{\partial {{n}_{0}}} \right) =
\frac{1}{m n_0} \left( n_0 \bar{s} \hbar \omega_c \right) =
\frac{2\pi {{\hbar }^{2}}}{{{m}}^{2}}{{n}_{0}} \frac{2 \bar{s}}{\nu}.
\end{align}
This result coincides with the former one for the integer quantum Hall case, where $\bar{s} = \nu/2$,~\cite{levay95,read09,read11,bradlyn12} so we will stick to Eq.~(\ref{eqn:s2_B0}) below.

Now we can calculate the dimensionless quantities:
\begin{align}
{{\tilde{\Omega }}}^{2}
\equiv \frac{{\Omega}_{q}^{2}}{sq{{\omega }_{c}}}
= \frac{\nu}{\sqrt{ 2\pi {n}_{0}a_{B}^{2} } },
\end{align}
where $a_B = \hbar^2\epsilon/m e^2$ is the Bohr radius in the material, and~\cite{read09,read11,bradlyn12}
\begin{align}
{{\tilde{\eta }}_{H}}
\equiv \frac{{{\omega }_{c}}{{\eta }_{H}}}{{{n}_{0}}{m}{{s}^{2}}}=
\frac{{{\omega }_{c}}}{{{n}_{0}}{m}{{s}^{2}}} \frac{\hbar n_0 \bar{s}}{2} =
\frac{{\bar{s}}}{2\nu}.
\end{align}

For a 2D electron gas (2DEG) at the GaAs/AlGaAs interface the effective mass (that is, the effective mass of the conduction band of GaAs) is ${m} \approx 0.067{{m}_{e}}$ ($m_e$ being the bare electron mass) and the dielectric constant is $\epsilon \approx 13$, so the Bohr radius is $a_B \approx 10$~nm. A typical number density is ${{n}_{0}} \sim {{10}^{13}}$~cm$^{-2}$. Thus, both ${{\tilde{\Omega }}}^{2}$ and $\tilde{\eta}_H$ are of order unity.
We will also define the dimensionless shear viscosity 
$\tilde{\eta}_{s} \equiv \omega_c \eta_s/n_0 m s^2$. It should be close to zero in the quantum Hall regime at low frequencies (long wavelengths), but we will also examine its effect when non-negligible.

\begin{figure} 
	\centering
	\includegraphics[width=\columnwidth]{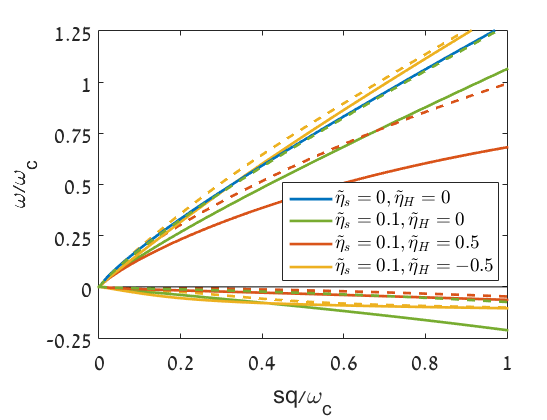}
	\caption{Dissipative magnetoplasmons: The edge magnetoplamon dispersion for different values of the dimensionless Hall and shear viscosities, $\tilde{\eta}_{s/H} \equiv \omega_c \eta_{s/H}/n_0 m s^2$. We took $\tilde{\Omega}=1$, $\tau \to \infty$, and $\zeta=0$. The continuous and dashed lines correspond to no-slip and no-stress boundary conditions, respectively. Different colors correspond to different values of the shear and Hall viscosities, as indicated in the legend. For $\eta_s>0$ the frequency has a negative imaginary part (decay), which is plotted with the same markings \emph{below} the horizontal axis.}
	\label{fig:dispersion_dissipation}
\end{figure}

\subsection{Edge megnetoplasmon dispersion}

We will now show some examples of the dispersions coming out of our results. We start from the dissipationless case ($\tau \to \infty$, $\zeta=\eta_s=0$) and display the edge magnetoplasmon dispersion in Fig.~\ref{fig:dispersion_dissipationless} for $\tilde{\Omega}=1$ for both the no-slip and no-stress boundary conditions.
For some values of the parameters the edge dispersion may terminate by overlapping with the bulk magnetoplasmon spectrum.
In Fig.~\ref{fig:dispersion_dissipation} we go on to examine the effect of finite shear viscosity on the dispersion, leaving the other parameters unchanged. The finite dissipation gives the dispersion a negative imaginary part, which is also plotted. Let us also note that the viscosities are expected to display frequency dependence~\cite{bradlyn12,gromov14,can14}, which can straightforwardly be incorporated into our formalism.

\begin{figure} 
	\centering
	\includegraphics[width=\columnwidth]{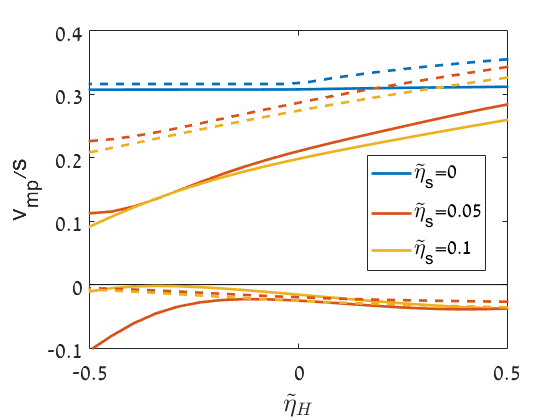}
	\caption{Coefficient of the leading term in the dispersion, $q\ln(\omega_c/s q)$ [see Eq.~(\ref{eq:fit})], as function of the dimensionless Hall viscosity, $\tilde{\eta}_{H} \equiv \omega_c \eta_H/n_0 m s^2$, for different values of the dimensionless shear viscosity (similarly defined). We took $\tilde{\Omega}=1$, $\tau \to \infty$, $\zeta=0$. The continuous and dashed lines correspond to no-slip and no-stress boundary conditions, respectively. For $\eta_s>0$ the coefficient has a negative imaginary part (decay), which is plotted with the same markings below the horizontal axis.}
	\label{fig:qlnq_coeff}
\end{figure}

Finally, we numerically examined the functional form of the dispersion at small $q$, and find that it is well described by a formula of the form
\begin{align} \label{eq:fit}
  \omega \sim v_{mp} q \ln \left( \frac{a \omega_c}{v q} \right),
\end{align}
with ``velocity'' $v_{mp}$ and dimensionless coefficients $a$. The leading term gives rise to magnetoplasmon velocity which diverges logarithmically as the wavevector goes to zero, a result of the long-range Coulomb potential~\cite{volkov88,aleiner94,aleiner95} if the Fetter approximation~\cite{fetter85} is not employed. It is sensitive to the viscosity, as shown in Fig.~\ref{fig:qlnq_coeff}. This is in contrast to the bulk dispersion, Eq.~(\ref{eq: bulk}), where viscosity effect enter only to order $q^2$.~\cite{tokatly07,tokatly09} This behavior is a result of the finite decay length of the edge modes into the bulk, which implies that even for small $q$ the electron density and velocity have non-negligible 2D Fourier components.
 

\section{Conclusions} \label{sec:conclusions}
To conclude, we have investigated the effect of viscosity (both the standard dissipative bulk and shear viscosity, as well as the symmetry-protected topological, non-dissipative Hall viscosity) on the dispersion of edge magnetoplasmons in 2D electronic systems under the influence of a magnetic field. Extending the Wiener-Hopf technique we have been able to find a general solution to the problem, which can be applied for arbitrary boundary conditions or frequency dependence of the viscosities. As opposed to bulk magnetoplasmons, whose dispersion is only affected by viscosity starting at second order in the wavevector~\cite{tokatly07,tokatly09}, on the edge the dispersion is already affected by the viscosity to lowest order in the wavevector, making it easier to extract experimentally.
Our calculations indicate that the relevant frequencies ($\lesssim \omega_c$) are in the microwave to THz regime, the wavelengths ($\gtrsim s/\omega_c$) are in the tens of nm range and above, and the velocities ($\sim s$) are of order of $10^3$~km/s, in both semiconductor heterostructures and graphene. These are thus within reach of present edge magnetoplasmon measurements~\cite{glattli85,volkov86,andrei88,grodnensky90,wassermeier90,ashoori92,talyanskii93,zhitenev93,sukhodub04,kamata10,kumada11,petkovic13,sasaki16,bosco17,endo18,jin18}, or possibly adaptations of bulk techniques~\cite{batke85,sohn1995dispersive,kukushkin2009dispersion,bandurin18}.
In the future it would be interesting to use the tools developed here to study the effects of viscosity on the analogous modes driven by nonzero Berry curvature instead of magnetic field~\cite{song16,jin16,principi16,principi17,zhang17,jia17,mahoney17,zhang18}.

\begin{acknowledgments}
We would like to thank L.I.~Glazman for suggesting to us to investigate this problem. 
Support by the Israel Science Foundation (Grant No.~227/15), the German Israeli Foundation (Grant
No.~I-1259-303.10), the US-Israel Binational Science Foundation (Grant No.~2014262), and the Israel Ministry of Science and Technology (Contract No.~3-12419), is gratefully acknowledged.
\end{acknowledgments}
	
\appendix
\section{The Fetter approximation} \label{sec:fetter}

The need resort to the Wiener-Hopf method arose since the solution of Poisson equation led to a potential-density relation, Eq.~(\ref{eq: potential Fourier}), with a kernel $\bar{L}(k)$ which is not a meromorphic function;
As a result, one cannot write down a local 2D partial differential equation relating the potential and the charge density.
To simplify the problem, Fetter~\cite{fetter85} suggested to replace $\bar{L}(k)$ by a meromorphic function, $\bar{L}_0(k)$, which agrees with $\bar{L}(k)$ to second order in $k$. It is given by
\begin{align}
\bar{L}_0 (k) = \frac{q}{k^2+2q^2} .
\label{eq:L_approximation}
\end{align}
In real space, this amounts to replacing the exact kernel, 
\begin{equation}
L(x)=
\frac{1}{2\pi } {{K}_{0}}(\left| q x \right|),
\label{eq:Lx_definition}
\end{equation}
with $K_0$ the modified Bessel function~\cite{gradshteyn}, by 
\begin{equation}
{{L}_{0}}(x)={{2}^{-3/2}}{{e}^{-\sqrt{2}q\left| x \right|}}.
\end{equation}

With the Fetter approximation, $-L_0/q$ is the Green function of the operator $\partial_x^2-2q^2$, allowing one to replace the 3D Poisson equation~(\ref{eq: Poisson}) by the 2D local linear equation
\begin{align}
  \left( \partial_x^2 - 2q^2 \right) \phi(x) = \frac{4\pi e}{\epsilon} q \delta n(x).
\end{align}
Combining this with the 2D local linearized continuity and Navier-Stokes equations~(\ref{eq: continuity})-(\ref{eq: NS}), one can seek for a solution which is a combination of exponential functions of $x$, and impose the boundary conditions. This readily leads to the dispersion relation.

One may also specialize our general solution, Eqs.~(\ref{eq: dispertion}), (\ref{eq: kh definition}), and~(\ref{eq: X definition}), to the Fetter approximation, by plugging in the kernel $\bar{L}_0$.
In this case the Riemann-Hilbert problem whose solution defines $\bar{X}_\pm$ [see Eq.~(\ref{eq: X definition})] involves rational functions only, and can be easily solved as
\begin{align}
{{\bar{X}}_{\pm }}(k)={{\left[ \frac{\left( k\pm {{k}_{H1}} \right)\left( k\pm {{k}_{H2}} \right)\left( k\pm i\sqrt{2}\left| q \right| \right)}{\left( k\pm {{z}_{1}} \right)\left( k\pm {{z}_{2}} \right)\left( k\pm {{z}_{3}} \right)} \right]}^{\pm 1}},
\label{eq:X_def_with_Fetters_apx}
\end{align}
where $z_{1,2,3}$ are the roots in the upper complex $k$ plane of the a sextic polynomial equation,
\begin{align}
(k^2+2q^2) &
\left[q{{s}_{\zeta }}^{2}{{\omega }_{s}(k)}\left( k^{2}+{{q}^{2}} \right)+q\omega \left( {\omega}^2_{H}(k)-{\omega}^2_{s}(k) \right) \right]
\nonumber \\
&+ 2{\Omega }_{q}^{2}{{\omega}_{s}(k)}\left( k^{ 2} + {q}^{2} \right) = 0.
\end{align}
We have verified numerically that the Wiener-Hopf dispersion, Eqs.~(\ref{eq: dispertion}), indeed agrees with the direct solution outline above in this case.


\section{Sufficiency of Eq.~(\ref{eq: determinant equations})} \label{sec:consistency}

In this Appendix we will verify that, for the no-slip boundary conditions, besides Eq.~(\ref{eq: determinant equations}), there are no additional equations to be satisfied in order for our solution to have the required analyticity properties and boundary conditions .

First, let us check that $\bar{\psi}_+$ is analytic in the upper complex plane. 
Performing the integral in Eq.~(\ref{eq: psi definition}) gives
\begin{widetext}
\begin{align}
\begin{split}
\bar{\psi}_+(k) = 
\frac{1}{\bar{X}_+(k)}\frac{1}{{{k}^{2}}+{{q}^{2}}}\left[ Ak+iqB+\left( iqA+kB \right)\frac{{{\omega }_{H}}}{{{\omega }_{s}}} \right]
+\frac{1}{2\left| q \right|}\left( \frac{{{C}_{-}}}{k+i\left| q \right|}-\frac{{{C}_{+}}}{k-i\left| q \right|} \right)+\frac{1}{2{{k}_{s}}}\frac{{{\omega }_{H}}({{k}_{s}})}{{\tilde{\omega }}}\left( \frac{{{D}_{+}}}{k-{{k}_{s}}}-\frac{{{D}_{-}}}{k+{{k}_{s}}} \right).
\end{split}
\end{align}
This expression has two potential poles in the upper complex $k$ plane, at $k=i|q|$ and $k=k_s$. The corresponding residues are 
\begin{subequations}
\begin{align}
\mathrm{Res} \bar{\psi}_+ (i|q|) & =
\frac{1}{{{\bar{X}}_{+}}(i\left| q \right|)}\frac{1}{2i\left| q \right|}\left[ Ai\left| q \right|+iqB+\left( iqA+i\left| q \right|B \right)\frac{{{\omega }_{H}}(i\left| q \right|)}{{{\omega }_{s}}(i\left| q \right|)} \right]-\frac{1}{2\left| q \right|}{{C}_{+}}, \\
\mathrm{Res} \bar{\psi}_+ (k_s) & = \frac{1}{{{\bar{X}}_{+}}({{k}_{s}})}\frac{1}{{{k}_{s}}^{2}+{{q}^{2}}}\left( iqA+{{k}_{s}}B \right)\frac{{{\omega }_{H}}({{k}_{s}})}{2i{{\gamma }_{s}}{{k}_{s}}}+\frac{1}{2{{k}_{s}}}\frac{{{\omega }_{H}}({{k}_{s}})}{{\tilde{\omega }}}{{D}_{+}}.
\end{align}
\end{subequations}
\end{widetext}
Both can be shown to vanish when substituting the definitions of $C_+$ and $D_+$. This, together with the corresponding behavior of $\bar{\psi}_-$, is sufficient to guarantee the correct analyticity properties of the Fourier-transformed potential $\bar{\phi}_\pm$.

Let us now turn to the Fourier-transformed velocities. 
From Eqs.~(\ref{eq set: positive Fouriers}) we obtain
\begin{subequations}
\begin{align}
\bar{v}_{x}^{+}&=\frac{1}{k}\left( \omega \frac{{{{\overline{\delta n}}}_{+}}}{{{n}_{0}}}-q\bar{v}_{y}^{+} \right), \\
\bar{v}_{y}^{+}&=\frac{1}{q{{\omega }_{H}}-ik{{\omega }_{s}}}\left[ iqk\frac{e}{m}{{\overline{\phi }}_{+}}-ikB+\left( \omega {{\omega }_{H}}-iqk{{s}_{\zeta }}^{2} \right)\frac{{{{\overline{\delta n}}}_{+}}}{{{n}_{0}}} \right].
\end{align}
\end{subequations}
The correct behavior of $\bar{v}_x^+$ is thus determined by that of $\bar{v}_y^+$. For the latter it remains to check the residue at
${{\omega }_{H}}=ik{{\omega }_{s}}/q$.
Using 
Eq.~(\ref{eq: n_bar}) for $\overline{\delta n}_+/n_0$ it can be shown to vanish. 

Finally, for any function $f(x)$ defined for $x<0$, with a corresponding Fourier transform $\bar{f}_+(k)$ analytic in the upper complex plain and vanishing as $|k| \to \infty$, its real-space boundary values are given by $f(0) = -i \lim_{|k|\to \infty} k \bar{f}(k)$, and $\partial_x f(0) = \lim_{|k|\to \infty} k^2 \bar{f}(k)$ if $f(0)=0$.
Since the Fourier functions are analytic in the upper complex plane, the boundary conditions for the velocity are obeyed automatically.
Using this it is straightforward to verify that all the boundary conditions are obeyed by our solution. For example, from the leading behavior of $\bar{v}_y^+$ at large $|k|$,
\begin{align*}
\bar{v}_{y}^{+}(k) \sim -\frac{1}{{{\gamma }_{s}}{{k}^{3}}}\left[ ikB+\omega {{\gamma }_{H}}\frac{i{{\gamma }_{s}}\frac{e}{m}{{k}^{2}}{{{\bar{\phi }}}_{+}}(\infty )-k{{\gamma }_{s}}A+ik{{\gamma }_{H}}B}{i{{\gamma }_{s}}{{s}_{\zeta }}^{2}+\omega \left( {{\gamma }_{H}}^{2}+{{\gamma }_{s}}^{2} \right)} \right] , 
\end{align*}
one can verify that $v_y(0)=0$ and $\partial_y v_y(0)$ equals the value encoded in $A$ and $B$ [cf.\ Eq.~(\ref{eq: phi+ explicit})]. Similar relations hold for all the other physical quantities, without imposing any additional constraints. The only exception is the potential for the no-stress boundary conditions, as discussed in Subsection~\ref{subsec:nostress}.


\bibliographystyle{apsrev4-1}
\bibliography{viscedge_refs}	

\end{document}